\documentclass[twocolumn,secnumarabic,amssymb,preprint nobibnotes, aps, prb]{revtex4-2}
\pdfoutput=1
\usepackage{graphics,dcolumn,float}
\usepackage{graphicx}
\usepackage{dcolumn}
\usepackage{bm}
\usepackage{amsmath}
\usepackage{amssymb}
\usepackage{color}
\begin{document}
\preprint{APS/123-QED}
\title{Spin Hall and Edelstein Effects in pseudospin-1 systems: significant contribution of vertex corrections}

\author{S. Rastegar}
\email[Corresponding author's Email: ]{s.rastegar@azaruniv.ac.ir}
\affiliation{Department of Physics,
Azarbaijan Shahid Madani University, Tabriz 53714-161, Iran}
\affiliation{Molecular Simulation Laboratory (MSL), Azarbaijan Shahid Madani University, Tabriz, Iran}
\author{A. Phirouznia}
\affiliation{Department of Physics,
Azarbaijan Shahid Madani University, Tabriz 53714-161, Iran}
\affiliation{Condensed Matter Computational Research Lab.,
Azarbaijan Shahid Madani University, Tabriz 53714-161, Iran}

\date{\today}
\begin{abstract}
Edelstein and spin Hall effect response functions for a two-dimensional (2D) system of pseudospin-1 particles is investigated. These two response functions denoted by $ \sigma_{SH} $ and $ \sigma_{EE} $ have been analyzed in a pseudospin-1 particle system in the presence of the Rashba-type spin-orbit interaction using the Kubo-Streda technique and vertex corrections with non-magnetic impurities. Then, for a given range of the Rashba coupling, response functions of the spin Hall effect (SHE) and Edelstein effect (EE) have been estimated at various energy gaps and Fermi energies. Results indicate that in this type of the two-dimensional materials, SHE and EE are essentially induced by the vertex corrections and without considering these corrections the amount of these effects are really negligible. It has also been realized that SHE and EE conductivities can be modulated by the Rashba coupling strength at low and intermediate level of this spin-orbit type interaction.
\end{abstract}


\maketitle

\section{INTRODUCTION}

\begin{figure}[h]
\includegraphics[width=0.8\linewidth]{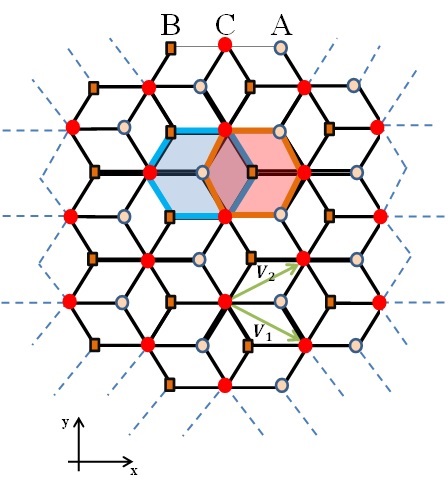}
\centering
\caption{A schematic representation of the $ \mathcal{T}_{3}$  lattice for pseudospin-1 fermion sample in the $ xy $ plane. Translation vectors are $ \mathbf{V}_{1} = (3/2;-\sqrt{3}/2)a $ and $ \mathbf{V}_{2} = (3/2;\sqrt{3}/2)a $, with lattice constant $ a $. Sites A and B that showed by solid circles and squares, making a hexagonal lattice, respectively. whereas, solid circles mark the hub sites C making a triangular lattice. \label{Fig-1.jpg}}
\end{figure}

Pseudospin-1 particle compounds have recently attracted a lot of interest in condensed matter physics. Low-energy gapless pseudospin-1 fermions have an energy spectrum that is linear in momentum, except for a flat band at zero energy, just like electrons in graphene. One of the most prevalent two-dimensional realizations of pseudospin-1 fermions (2D) is the $ \mathcal{T}_{3}$  lattice. Atoms at the centers and vertices of the hexagonal lattice could be used to create a simple tight-binding model for this type of materials. Due to the placement of three sites per unit cell, the electron states in this model are given by three-component fermions with three bands of the energy spectrum. Such a system is shown schematically in Fig. \ref{Fig-1.jpg}.

The spin Hall effect (SHE) has received much attention because  of its prospective applications in spintronic technologies. Hirsch \cite{hirsch1999spin} explored the spin Hall effect (SHE), which Dyakonov and Perel \cite{d1971possibility,dyakonov1971current} studied in 1971. The spin Hall effect (SHE) and the inverse spin Hall effect (ISHE) \cite{sinova2015spin}, a group of transport phenomena \cite{saitoh2006conversion,valenzuela2006direct}, greatly aided the development of experimental spintronic devices \cite{dyakonov1971current,hirsch1999spin,zhang2000spin}. One may list examples of its use in memory, logic, and sensing devices. Kane and Mele discovered the quantum spin Hall effect in graphene in 2005 \cite{kane2005quantum}. The low-energy electronic structure of a single layer graphene with spin-orbit interactions was investigated. They found that graphene changes from its ideal two-dimensional semimetallic state at low temperatures to a quantum spin Hall insulator due to the symmetry-enabled spin-orbit potential. 

\begin{figure}[h]
\includegraphics[width=0.8\linewidth]{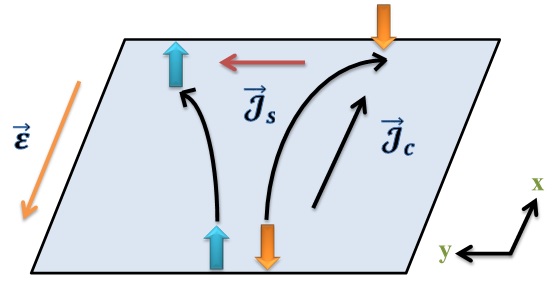}
\centering
\caption{Schematic representation of the SHE in a two-dimensional material. Both the charge and spin Hall currents are driven by an external electrical field. \label{Fig-2.jpg}}
\end{figure}

After initial findings about these phenomena in semiconductors and metals, it was known that the spin Hall effect might offer a new strategy for the interconversion of spin and charge indications \cite{kato2004observation,wunderlich2005experimental,saitoh2006conversion,valenzuela2006direct}. SHE, which results from strong spin-orbit coupling (SOC), may pave the way for generating pure spin current from charge current. In the spin Hall regime, a spin current is generated perpendicular to an applied electric field. Additionally, unlike SHE within the inverse SHE regime, spin currents can be converted into electric signals \cite{mosendz2010detection,azevedo2015azevedo}. 

Meanwhile, spin Seebeck effect \cite{uchida2008k,uchida2011long,qu2013intrinsic} demonstrates that the temperature gradient can generate spin current which confirms that the spin-orbit interactions are also crucial in the field of "spin caloritronics."  

The strength of this effect is given by spin Hall angle $ \gamma $ that measures the material's effectiveness according to relation $ \gamma = \mathbf{\mathcal{J}}_{\bot}/J_{\parallel}$, which is defined as the ratio between steady-state longitudinal charge-current, $J_{\parallel}$ and $ z $-polarized transverse spin currents given by $\mathbf{\mathcal{J}}_{\bot}$  as shown in Fig. \ref{Fig-2.jpg}. The material's SHE efficiency is determined by the spin Hall angle. The charge efficiency in the present spin transition can be determined using materials with a large spin Hall angle (SHA). 

The anomalous effect is another well-known phenomenon in this field. The nonmagnetic spin Hall effect \cite{nagaosa2010anomalous} and the well-known anomalous Hall effect (AHE) \cite{nagaosa2010anomalous} in ferromagnetic metals are closely linked. The magnetic field $ H_{z} $ and magnetization $ M_{z} $ both affect the Hall resistance $\rho_{H} $, through the relation $\rho_{H}=R_{0}H_{z}+R_{s}M_{z}$\cite{nagaosa2010anomalous}. Where, the normal and anomalous Hall resistances are $ R_{0} $ and $ R_{s} $, respectively \cite{pugh1930hall,pugh1932hall}. The anomalous Hall effect has become one of the most urgent issues in solid-state physics because it is challenging to forecast the carrier density in ferromagnets \cite{smit1955spontaneous,berger1970side,kondo1962anomalous,nozieres1973simple}.

The Edelstein effect (EE) \cite{edelstein1990spin} and the inverse Edelstein effects (IEE) \cite{mahfouzi2014spin,shen2014microscopic} are two additional effects that have recently attracted a lot of interest. In the regime of EE a continuous non-equilibrium spin polarization $ S_{y} $ is produced by a constant current $ J_{x} $ driven by an electric field $ E_{x} $. It was also demonstrated that the EE and the SHE are connected \cite{edelstein1990spin}. The inverse Edelstein effect is the process underlying the spin-to-charge transformation. The Edelstein effect is a promising phenomena for spintronics applications since it can produce spin polarization in nonmagnetic materials just electrically. Therefore, it is highly desirable to find materials that are highly efficient at "converting" the electric current into spin polarization. Meanwhile, it should be noted that the spin and charge quantum numbers can also be converted to pseudospin index in graphene-like materials \cite{Baghran_2021_1,Baghran_2021_2}.

Edelstein effect of enormous topological-insulator-graphene heterostructures has been investigated by Rodriguez-Vega et al.   \cite{rodriguez2017giant}, where they show that the nonequilibrium uniform spin-density accumulation caused by a charge current in magnetic TI-graphene heterostructures can be $ 10-100 $ times larger than in TIs alone, resulting in a massive Edelstein effect. 

Valley Edelstein effect (VEE) has been discovered in the monolayer transition-metal dichalcogenides by Taguchi et al \cite{taguchi2018valley}. They found that in gated monolayer transition-metal dichalcogenides (MTMDs), Berry curvatures resulting from coexisting Rashba and Ising SOCs combined with traditional Edelstein effects lead to valley-contrasting spin polarization parallel to the applied electric field. 

The ability to introduce SOC with multiple symmetries \cite{mccann2012z,pachoud2014scattering} and varying spatial extent is a fascinating property of two-dimensional materials. Recent theoretical investigations have clarified the significant role that the SOC symmetry plays in the resonant scattering domain \cite{pachoud2014scattering,yang2016extrinsic}. 

There is a microscopic approach for the spin Hall effect in amorphous materials using vertex corrections. This approach has been employed in amorphous graphene \cite{milletari2016quantum} where the nonperturbative quantum diagrammatic method which has been applied in this approach can also be used to obtain the response function in materials resembling graphene \cite{milletari2016quantum}.

For non-magnetic impurities, the vertex coefficient vanishes when the Fermi level is in the upper band (both bands are occupied). This problem leads to the disappearance of the spin-Hall effect in the thermodynamic limit for any degree of disorder \cite{nitta1997gate}. It has also been shown that the bare vertex is a good approximation and that the bare bubble diagram is adequate for calculating the spin-Hall conductivity in the presence of magnetic impurities \cite{moca2007spin}.

The SHA rises linearly with the SOC impurity density in disordered samples where other variables restrict charge mobility \cite{milletari2016crossover}.
The spin Hall current in graphene identically vanishes when the spin-orbit interaction (SOI) is absent even in the presence of the magnetic impurities \cite{ding2009charge}. Therefore, it can be inferred that the SOI is of crucial importance for SHE and it cannot be generated merely by the spin-flip scatterings \cite{ding2009charge}.
It has been shown that the spin Hall conductivity (SHC) are generally finite in the presence of SOI and magnetic impurities under a zero external magnetic field \cite{ding2009charge}.

When the magnetic impurities are combined with the SOI, the magnetic scattering centers act as spin-dependent barriers, causing a charge imbalance at the boundaries. Changing the chemical potential close to the gap should also reveal charge and spin Hall effects in graphene. Several investigations have been performed on the charge and spin Hall phenomena in magnetically impure graphene. Using the Kubo formula, analytical formulations of the charge and spin Hall conductivities has been developed theoretically. \cite{ding2009charge}.
 
The SHE and EE can be described by the 	Edelstein conductivity (EC) and the SHC that are defined through the following relations $ J_{y}^{z} = \sigma^{SHE}_{xy}E_{x} $, $ s^{y} = \sigma^{EE}_{xy}E_{x} $ respectively. In this study spin Hall and Edelstein conductivities for pseudospin-1 particles with vertex corrections have been investigated. Using the Kubo-Streda method and vertex corrections in the presence of non-magnetic impurities, these two response functions, $ \sigma^{SHE} $ and $ \sigma^{EE}$, have been evaluated in a pseudospin-1 particle system with Rashba-type spin-orbit interaction. Numerical calculations provide the normalized operators within a self-consistent approach. Results show that the bare Kubo response function of both EE and SHE are absolutely small and the main contribution comes the vertex corrections. Therefore, in these type of materials we cannot expect high SHA. Meanwhile, EC and SHC could reveal the different physics behind these novel materials.

\section{METHODOLOGY}
Within a low energy approximation, effective Hamiltonian of massive Dirac-like pseudospin-1 particles in the presence of the Rashba spin-orbit interaction can be stated as follows: 
\begin{align}
H &= \hbar v_{F} (S_{x}k_{x}\otimes I_{\sigma} + S_{y}k_{y}\otimes I_{\sigma})
\nonumber
\\
&+\lambda_{R}(k_{x}I_{\delta}\otimes\sigma_{y}-k_{y}I_{\delta}\otimes \sigma_{x}) + m S_{z} \otimes I_{\sigma}
\end{align}
where $ v_{F} $ is the pseudospin-1 particle's Fermi velocity, $ \lambda _{R} $ is the Rashba coefficient and $ \mathbf{S}=(S_{x},S_{y},S_{z}) $ is a vector of matrices with pseudospin components \cite{bercioux2009massless}: 
\begin{equation*}
\mathbf{S_{x}}=\frac{1}{\sqrt{2}} \left(\begin{array}{ccc}
0 & 1 & 0\\
1 & 0 & 1\\
0 & 1 & 0
\end{array}\right), \quad 
\mathbf{S_{y}}=\frac{1}{\sqrt{2}} \left(\begin{array}{cccc}
0 & -i & 0\\
i & 0 & -i\\
0 & i & 0
\end{array}\right),
\end{equation*}\\
\begin{equation*}
\mathbf{S_{z}}=\left(\begin{array}{ccccc}
1 & 0 & 0\\
0 & 0 & 0\\
0 & 0 & -1
\end{array} \right).
\end{equation*}\\
The three matrices represent all pseudospin-1 particles that fulfill the angular momentum commutation relations $ [S_{l},S_{m}]=i\epsilon_{lmn}S_{n}$ with three eigenvalues, $ s=\pm 1, 0 $, where $ \epsilon_{lmn} $ is the Levi-Civita symbol. However, they do not form a Clifford algebra, as opposed to Pauli matrices; that is, $ \{S_{n},S_{m}\} \neq 2 \delta_{n,m} I_{3\times 3} $. These three dimensional matrices represent the contribution of three A, B and C sublattices.
\\
\\
After the Fourier transformation
\begin{equation}
\mathcal{H} = \sum_{\mathbf{k}}\Psi_{\mathbf{k}}^{\dagger} H(\mathbf{k})\Psi_{\mathbf{k}},
\end{equation}
where $ \Psi_{\mathbf{k}}^{\dagger} = (c_{A,\mathbf{k},\uparrow}, c_{B,\mathbf{k},\uparrow}, c_{C,\mathbf{k},\uparrow}, c_{A,\mathbf{k},\downarrow}, c_{B,\mathbf{k},\downarrow}, c_{C,\mathbf{k},\downarrow}) $ and $ c_{ik\sigma}^{\dagger} $ denotes the creation operator of electron on the $ i $ sublattice with spin of $ \sigma $ and wave-number of $ k $.
\\
\\
Then in presence of Rashba coupling, the Hamiltonian is given by
\begin{align}
H &= \frac{\hbar v_{F}}{\sqrt{2}} \left(\begin{array}{ccc}
\nonumber
0 & k_{-} & 0\\
k_{+} & 0 & k_{-}\\
0 & k_{+} & 0
\end{array}\right)\otimes I_{\sigma}\\
\nonumber
&\quad + \lambda _{R} \left(\begin{array}{ccc}
k_{x} & 0 & 0\\
0 & k_{x} & 0\\
0 & 0 & k_{x}
\end{array}\right)\otimes \sigma _{y} - \lambda_{R} \left(\begin{array}{ccc}
k_{y} & 0 & 0\\
0 & k_{y} & 0\\
0 & 0 & k_{y}
\end{array}\right)\otimes \sigma _{x}\\
&\quad + m \left(\begin{array}{ccc}
1 & 0 & 0\\
0 & 0 & 0\\
0 & 0 & -1
\end{array}\right)\otimes I_{\sigma}
\end{align}
The energy spectrum can be obtained as
\begin{align}
E_{1,2}(\mathbf{k}) &= \pm \arrowvert \mathbf{k} \arrowvert \lambda_{R}
\nonumber
\\
E_{3,4}(\mathbf{k}) &= \pm \sqrt{m^{\prime}(\mathbf{k}) + k^{2} \lambda_{R}^{2}-2\sqrt{k^{2}\lambda_{R}^{2}m^{\prime}(\mathbf{k})}}
\nonumber
\\
E_{5,6}(\mathbf{k}) &= \pm \sqrt{m^{\prime}(\mathbf{k}) + k^{2} \lambda_{R}^{2}+2\sqrt{k^{2}\lambda_{R}^{2}m^{\prime}(\mathbf{k})}},
\label{6}
\end{align}
where $ m^{\prime}(\mathbf{k}) = m^{2} + k^{2}\hbar^{2} v_{F}^{2} $.

Therefore, electrons in the $ \mathcal{T}_{3} $ lattice as seen in Fig. \ref{Fig-1.jpg},   can behave as massless Dirac fermions with pseudospin $ S = 1 $, rather than $ S = 1/2 $ for those trapped in the hexagonal lattice. Each unit cell in this $ \mathcal{T}_{3} $ lattice has three inequivalent sublattices. Two lattice sites, A and B, are triply coordinated, while site C connects six of its closest neighbors. 

\section{The Kubo-Streda approach}
The density operators for charge and spin-current can be defined as follows \cite{rashba2003spin,zhang2005intrinsic,zhang2008theory}:
\begin{equation}
\mathbf{J}=-e\Psi^{\dagger}(\mathbf{x})\mathbf{v}\Psi(\mathbf{x}),
\end{equation}
\begin{equation}
\mathbf{\mathcal{J}}_{\beta}^{\alpha}=-e\Psi^{\dagger}(\mathbf{x})\frac{1}{2}\{s_{\alpha},\mathbf{v}_{\beta}\}\Psi(\mathbf{x}),
\end{equation}
where $ \{.,.\} $ represents the anticommutator, $ \mathbf{v} $ represents the velocity operator, $ -e $ represents the electron$ ^{,} $s charge, and $ s_{z}\equiv s_{3} $ represents the diagonal Pauli matrix with eigenvalues $ \pm 1 $. Using the Kubo-Streda method the conductivity tensor in the presence of SOC \cite{crepieux2001crepieux} is:
\begin{equation}
\sigma_{ij}^{z}=\frac{\hbar}{2\pi \Omega} \mathrm{Tr} \big<J_{i} (G^{R}-G^{A})\mathcal{J}_{i} G^{A} - \mathcal{J}_{j} (G^{R}-G^{A})J_{i}G^{R}\big>_{dis},
\end{equation}
$ G^{R(A)} $ denotes the retarded (advanced) Green$ ^{,}s  $ function, which is defined as follows: 
\begin{equation}
G^{R(A)}=\frac{1}{\epsilon - H_{0}-V\pm i0^{+}},
\end{equation}
Where $ H $ stands for the total Hamiltonian, $ H=H_{0}+V $, with $ H_{0} $ for the non-perturbed Hamiltonian, and $ V $ represents the perturbation. Cartesian elements of the charge- and spin-current density operators' are denoted by $ J_{i} $ and $ \mathcal{J}_{i} $ respectively. $ \big<...\big>_{dis} $ indicates the configurational disorder average compared to the conventional version. 
\begin{equation}
\big<O\big>_{dis}=\lim\limits_{N,\Omega \rightarrow \infty} \bigg(\prod_{i=1}^{N}\int\limits_{\Omega} \frac{d^{2}x_{i}}{\Omega}\bigg)O(\mathbf{x}_{1},...,\mathbf{x}_{N})\bigg|_{\frac{N}{\Omega}=n},
\end{equation}
Where $ i $ represents impurities, $ \Omega $ represents the sample area, and Tr represents the trace over the entire Hilbert space. The conductivity of SH is equal to 
\begin{equation}
\sigma^{SHE}_{xy}=\frac{1}{\pi \Omega} \mathrm{Tr}[\big<G^{R} J_{x} G^{A}\big>_{dis}\mathcal{J}_{y}]
\end{equation}
and also EE conductivity is given as follows
\begin{equation}
\sigma^{EE}_{xy}=\frac{1}{\pi \Omega} \mathrm{Tr}[\big<G^{R} S_{y} G^{A}\big>_{dis}J_{x}]
\end{equation}
For short-range impurities, calculations were made using the weak-scattering regime. Within the Kubo formalism, the main focus has been devoted on analyzing the electric dc Edelstein and Hall conductivities in a typical pseudospin-1 particle system with nonmagnetic disorders. 

The linear response Kubo formalism has been used to explain the fully quantum mechanical response functions by taking the vertex correction into account. The response functions were obtained at zero temperature, which may be specified using Green's time-ordered functions. Because the calculations were performed in the dc regime, it can be achieved by $ \omega \longrightarrow 0 $ limit at the end of the calculations. In the meantime, this method has been necessary for all ac-based calculations in this approach. 

\subsubsection{Disordered Vertex Correction}
A randomly distributed disorder potential has been assumed as follows
\begin{equation}
\hat{V}(\mathbf{r}) = V_{0}\sum_{i=1}^{N}\delta(\mathbf{r}-\mathbf{R}_{i}),
\end{equation}
weak, and short-range Gaussian correlation can be considered by imposing following relations 
\begin{equation}
\big<\hat{V}(\mathbf{r})\big>_{imp} = 0,\quad \big<\hat{V}(\mathbf{r}_{1})\hat{V}(\mathbf{r}_{2})\big>_{imp} = nV_{0}^{2}\delta(\mathbf{r}_{1}-\mathbf{r}_{2}),
\end{equation}
The impurity concentration and scattering potential are indicated by $ n=5\times 10^{10}cm^{-2} $ and $ V_{0} =0.1 eV$, respectively. This method has been used extensively to study the properties of disordered systems. The Kubo relation for dc electric current conductivity is: 
\begin{equation}
\sigma_{x\nu} = \frac{\hbar}{2\pi L^{2}} \mathrm{Tr}\big<\hat{j}_{x}\hat{G}^{R}\hat{j}_{\nu}\hat{G}^{A}\big>_{imp},
\end{equation}
where $ L^{2} $ represents the system area and $ \hat{G}^{R(A)}(\epsilon) = (\epsilon \pm i0 - \hat{H} - \hat{V})^{-1} $ refers to  the retarded (advanced) Green$ ^{,} $s function in the Pauli space. The current operator is expressed as $ \hat{\mathbf{j}}=-e(-i/\hbar)[\hat{\mathbf{r}},\hat{H}]=-ev_{F}\hat{\mathbf{z}}\times \hat{\mathbf{\sigma}}. $ Meanwhile, $ \big<...\big>_{imp} $ shows an ensemble average over random nonmagnetic impurity configurations. The conductivity then is given as
\begin{align}
\sigma_{x\nu} &\approx \frac{\hbar}{2 \pi L^{2}} \mathrm{Tr}[\hat{j}_{x}\big<\hat{G}^{R}\big> \hat{j}_{\nu} \big<\hat{G}^{A}\big>] + \text{Vertex correction}
\\
\nonumber
&\equiv \frac{\hbar}{2 \pi L^{2}} \mathrm{Tr}[\hat{j}_{x}\big<\hat{G}^{R}\big> \hat{\tilde{j}}_{\nu} \big<\hat{G}^{A}\big>],
\end{align}
Where $ \big<\hat{G}^{R(A)}\big> $ is the averaged Green$ ^{,} $s function and $ \hat{{j}}_{\nu} $ is the bare charge current density operators. The ladder diagram correction to the bare charge current density operator can be evaluated using the following relation\cite{sinitsyn2007anomalous},
\begin{equation}
\hat{\tilde{j}}_{\nu}(\varepsilon_{F}) \equiv \hat{j}_{\nu} + \delta \hat{\tilde{j}}_{\nu}(\varepsilon_{F}),
\end{equation}
where $ \tilde{j}_{\nu} $ is the normalized charge current operator within the ladder vertex correction and $ \delta \hat{\tilde{j}}_{\nu} $ is the charge current vertex correction. The averaged Green function is expressed using the Dyson equation and the Born approximation, as shown in Fig. 3. 
\begin{align}
\big<\hat{G}^{R(A)}\big> &= \big<(z-\hat{H}-\hat{V})^{-1}\big>_{imp}
\\
\nonumber
&= \hat{G}_{0}^{R(A)}+\hat{G}_{0}^{R(A)}\hat{\varSigma}^{R(A)}\big<\hat{G}^{R(A)}\big>,
\end{align}
with $ z=\epsilon\pm i0 $ in which $ \epsilon $ is small positive. Here, the solution may then be written as:
\begin{equation}
\big<\hat{G}\big> = ((\hat{G}_{0}^{R(A)})^{-1} - \hat{\varSigma}^{R(A)})^{-1},
\end{equation}
in which the self-energy in the Born approximation is
\begin{equation}
\hat{\varSigma}^{R(A)} = \big<\hat{V}\big>_{imp} + \big<\hat{V}\hat{G}_{0}^{R(A)}\hat{V}\big>_{imp}, 
\end{equation}
in which the impurity configuration average can be obtained as 
\begin{equation}
 \big<\hat{V}\hat{G}_{0}^{R(A)}\hat{V}\big>_{imp} = nV_{0}^{2}\int\frac{d^{2}q}{(2\pi)^{2}}\hat{G}_{0}^{R(A)}. 
\end{equation}
The total of all ladder diagrams in Fig. 3 depicts the vertex function in the Born approximation. The ladder vertex-normalized charge current, $ \hat{j}_{\nu} $, is followed by the integral (Bethe-Salpeter) equation as shown in Fig. 3 which can be expressed as following relations: 
\begin{equation}
\hat{\tilde{j}}_{\nu} = \hat{j}_{\nu} + nV_{0}^{2} \int \frac{d^{2}\mathbf{q}}{(2 \pi)^{2}} \big<\hat{G}^{R}\big> \hat{\tilde{j}}_{\nu} \big<\hat{G}^{A}\big>. 
\end{equation}
\begin{align}
\sigma^{SHE}_{xy} & =\frac{1}{\pi \Omega} \mathrm{Tr}[\mathcal{G}^{R} J_{x} \mathcal{G}^{A} \mathcal{J}_{y}]
\nonumber\\
& + \frac{1}{\pi \Omega} \mathrm{Tr}[\mathcal{G}^{A} \mathcal{J}_{y} \mathcal{G}^{R}\big<\mathcal{T}^{R}\mathcal{G}^{R}J_{x}\mathcal{G}^{A}\mathcal{T}^{A}\big>_{dis}],
\end{align}
and similarly Bethe-Salpter equation for the EC results in:
\begin{align}
\sigma^{EC}_{xy} & =\frac{1}{\pi \Omega} \mathrm{Tr}[\mathcal{G}^{R} J_{x} \mathcal{G}^{A} \mathcal{S}_{y}]
\nonumber\\
& + \frac{1}{\pi \Omega} \mathrm{Tr}[\mathcal{G}^{A} \mathcal{S}_{y} \mathcal{G}^{R}\big<\mathcal{T}^{R}\mathcal{G}^{R}J_{x}\mathcal{G}^{A}\mathcal{T}^{A}\big>_{dis}].
\end{align} 
\begin{figure}[h]
\includegraphics[width=1\linewidth]{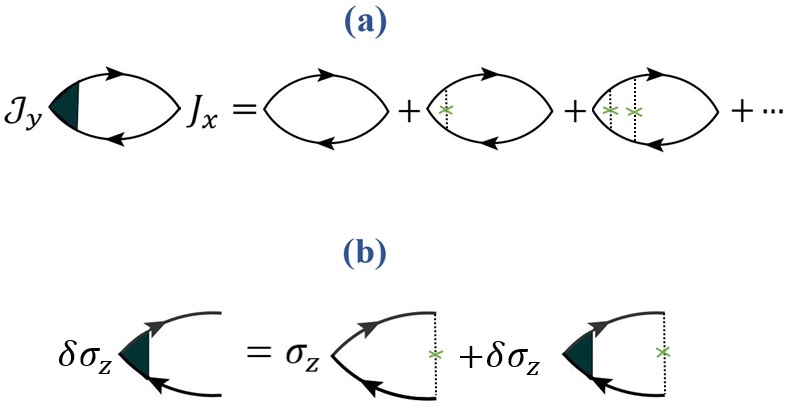}
\centering
\caption{(a) Ladder diagram of the response function successive interactions of an electron-hole pair with impurities. (b) Iterative equation of vertex corrections (Bethe-Salpeter equation) is presented using the diagrammatic approach. \label{Fig-3.jpg}}
\end{figure}
\section{Results and Discussion}

It can be realized that both the EE and SHE are obtained as a result of the vertex corrections, whereas in the absence of these corrections, the magnitudes of the EE and SHE response function are negligible. Accordingly, vertex correction plays a significant role in the realization of SHE and EE in pseudospin-1 systems. As a result, the EE and SHE in pseudospin-1 particle systems cannot be considered first order effects that could be captured by a linear response function or band broadening given by the relaxation time broadening of self-energies within the Born approximation. Unlike the pseudospin-1 particles, it has been shown that the spin Hall conductivity of two-dimensional electron gas systems with isotropic short-range defects identically vanishes with the vertex corrections \cite{inoue2004suppression}. This is due to the fact that the dressed current operator, which is the renormalized vertex-modified current, is corrected in such a way that the spin dependent part of the electric current operator, that comes from the Rashba coupling contribution in the current operator, is suppressed by the vertex corrections \cite{inoue2004suppression}. In a two-dimensional electron gas system, this means that vertex corrections decrease the correlation between electric current and spin current. This makes the SH conductivity go away. However, As mentioned before, vertex corrections could effectively generate accountable SH conductivity in pseudospin-1 systems.

In the absence of the vertex corrections, we have realized that the magnitude of SH and EE conductivities is negligible, while the vertex corrections enhance these conductivities significantly. Therefore, it can be inferred that there is a deep difference between the influence of the vertex corrections in the electron gas and pseudospin-1 systems. On the other hand, it has been reported that the vertex corrections enhance the SHE in graphene, this can be obtained within the Dirac cone approximation by taking into account chiral bands \cite{sinitsyn2006charge}.

\begin{figure}[h]
\includegraphics[width=0.9\linewidth]{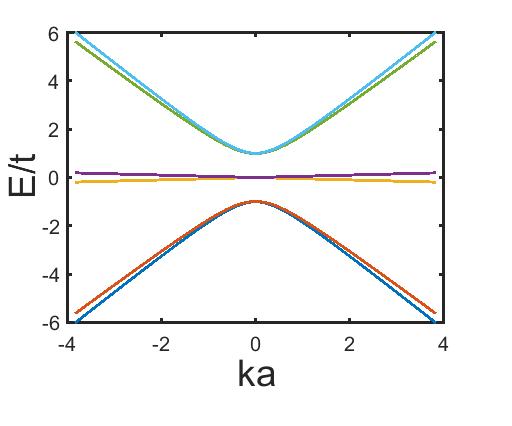}
\centering
\caption{Band structures of pseudospin-1 systems for $ m = 1.0 $, $ \lambda_{R} = 0.05$ and $ E_{f} = 0.25 $. \label{Fig-4.jpg}}
\end{figure}

Accordingly, this similar influence of the vertex corrections in graphene and pseudospin-1 systems suggests that the linear nature of the band energies in both the graphene and pseudospin-1 samples may provide this deep difference with the results of the electron gas systems. It should be noted that the pseudospin-1 systems with the Rashba spin-orbit coupling show linear dispersion as indicated by Eq. (2.5). However, due to the low Rashba coupling strength these linear bands cannot result in metallic sample within the Brillouin zone at $ m\neq0 $ as shown in Fig. \ref{Fig-4.jpg}.

\begin{figure*}[h]
\includegraphics[width=1.0\linewidth]{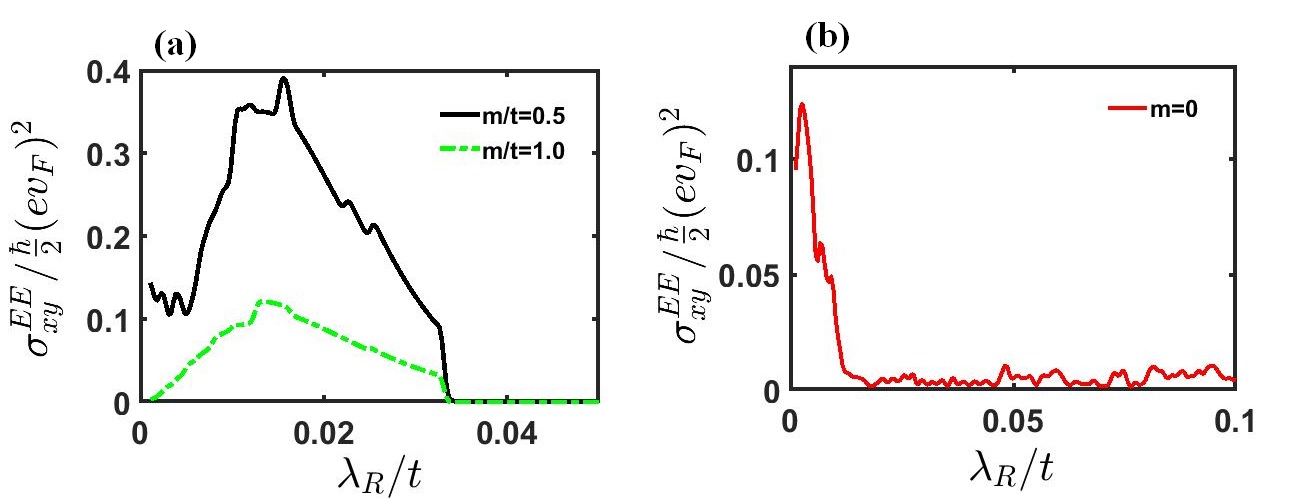}
\centering
\caption{The EE conductivity in (a) gapped systems for $ m = 0.5 $ and $ m = 1.0 $, and (b) gappless sample ($ m = 0 $). \label{Fig-5.jpg}} 
\end{figure*}

\begin{figure*}[h]
\includegraphics[width=1.0\linewidth]{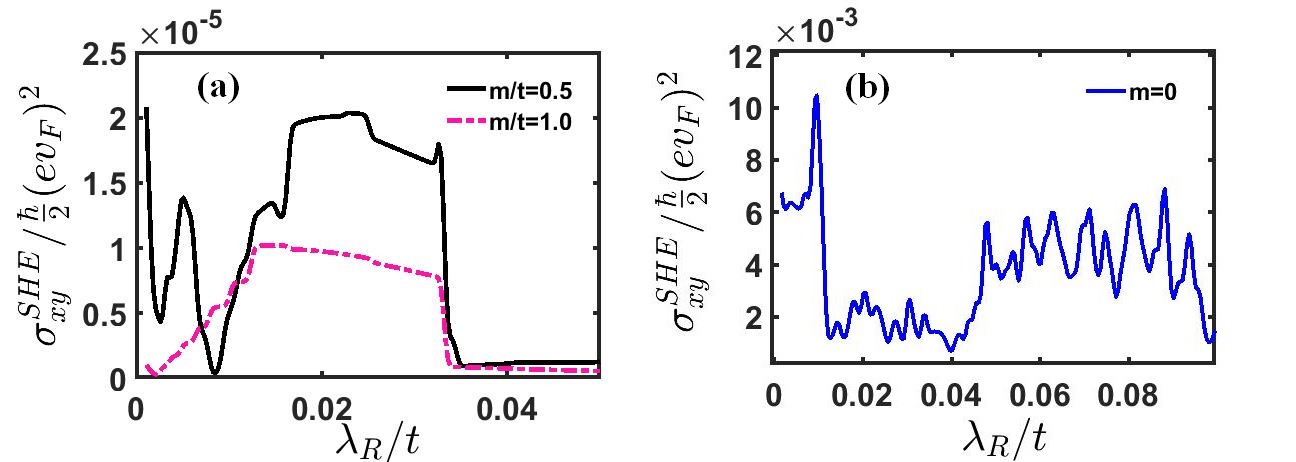}
\centering
\caption{The SHE conductivity in (a) gapped systems for $ m = 0.5 $ and $ m = 1.0 $, and (b) gappless sample ($ m = 0 $). \label{Fig-6.jpg}}
\end{figure*}

SHC and EC have been obtained as a function of the Rashba coupling strength at different gap values as shown in figures \ref{Fig-5.jpg} and \ref{Fig-6.jpg}. It can be seen that the SHC and EC can be manipulated by the Rashba coupling. It has also been realized that, at high Rashba couplings, the SHE and EE conductivities decrease rapidly. However, in the gapless pseudospin-1 systems, vertex corrections are very effective at low and intermediate Rashba couplings, where both SHE and EE conductivities are significantly increased as a result of the vertex corrections. At some especial Rashba coupling strength, SHE and EE conductivities abruptly fall in gapped pseudospin-1 systems.

As it can be verified numerically that scattering rate decreases by increasing the Rashba couplings. It seems that this simple picture could explain the suppression of the SHE and EE conductivities at high Rashba couplings. Meanwhile, it should be noted that the effect of the Rashba coupling should be considered beyond the first order linear response regime, since in the case of the pseudospin-1 systems, only the vertex corrections can contribute to the SHE and EE response functions, and as mentioned before, there is no first order contribution. Vertex correction takes into account the multiple independent scatterings in which both retarded and advanced branches of the electron propagator are interacting with impurities, as shown in Fig \ref{Fig-3.jpg}. Accordingly, we expect some different behavior when the single scatterings are dominant and the response function can be described within the first-order approximations. Physically, these multiple, consecutive scatterings that don't cross each other create a series of independent events called a ladder diagram. These diagrams represent the quantum side jump (QSJ) contribution to the SHE \cite{milletari2016quantum}.

Ladder diagrams should be used to explain the role of Rashba coupling in the framework of independent multiple scatterings. The ladder diagrams essentially measure the side-jump contribution, which corresponds to the transverse coordinate shift of the carriers after scattering process. Therefore, at the intermediate level of the Rashba coupling strength it seems that this shift is saturated \cite{sinitsyn2007anomalous} and further increasing of the Rashba coupling strength just randomizes the final spin state of the carriers, due to the increased spin-mixing provided by the increased Rashba coupling. Accordingly, the suppression of the EE and SHE is expected as a result of the Rashba coupling increment after this saturation limit. Meanwhile, there are other reports confirming that the SH conductivity of 2DEG is suppressed by scattering from short range nonmagnetic impurities with a linear Rashba interaction \cite{mishchenko2004spin,inoue2003diffuse,inoue2004suppression,raimondi2005spin,ol2005spin}.

As shown in Fig. \ref{Fig-5.jpg} and Fig. \ref{Fig-6.jpg}, EE is suppressed in the gapless sample, while SHE can be increased by an order of magnitude in that type of systems. Meanwhile, in gapped systems, it should be considered that both EE and SHE are increased by decreasing the gap value. This can be explained if we consider that increasing the gap value decreases spin-band mixing.

\begin{figure}[h]
\includegraphics[width=0.9\linewidth]{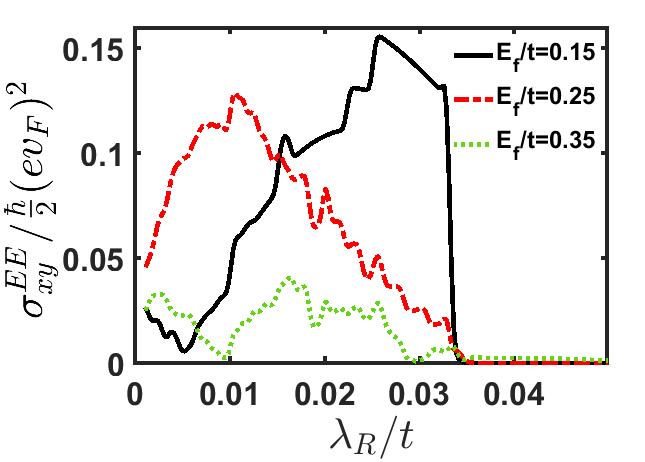}
\centering
\caption{The EE conductivity in systems with $ E_{f} = 0.15 $ (black line), $ E_{f} = 0.25 $ (red line) and $ E_{f} = 0.35 $ (green line). \label{Fig-7.jpg}}
\end{figure}

\begin{figure}[h]
\includegraphics[width=0.9\linewidth]{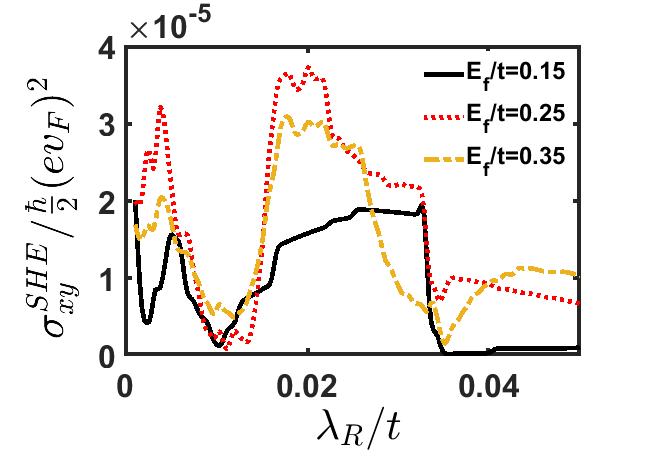}
\centering
\caption{The SHE conductivity in systems with $ E_{f} = 0.15 $ (black line), $ E_{f} = 0.25 $ (red line) and $ E_{f} = 0.35 $ (orange line). \label{Fig-8.jpg}}
\end{figure}

As shown in Fig. \ref{Fig-7.jpg} and Fig. \ref{Fig-8.jpg}, the SHE and EE do not have a monotonic dependence on Fermi energy. However, it has been mentioned that the SH conductivity in graphene is inversely proportional to the chemical potential \cite{ding2009charge}.


\section{CONCLUSIONS}
The spin Hall and Edelstein conductivities are studied numerically in the current work for the new type two-dimensional materials called pseudospin-1 systems. These two response functions have been evaluated in a pseudospin-1 particle system with Rashba-type interaction using the Kubo-Streda technique and vertex corrections with non-magnetic impurities. Numerical calculations were carried out to derive the normalized operators self-consistently. It has been shown that the vertex corrections play a significant role on the realization of the SHE and EE in the pseudospin-1 systems, where in the absence of these corrections SHE and EE response functions are essentially negligible. The ladder diagrams mainly capture the side jump effect contribution. Therefor, it can be inferred that the side jump effect is one of the main effects which can induce SHE and EE in pseudospin-1 systems. Meanwhile, the contribution of the vertex corrections completely removed at high Rashba couplings. When the Rashba coupling strength is high due to the saturation of the side jump shift of the carriers increased Rashba coupling just increases the spin relaxation that decrease the SHE and EE abruptly.
\bibliographystyle{apsrev4-1}
\bibliography{refrences.bib}
\end{document}